\documentclass[preprints,journal,article,accept,pdftex,moreauthors]{Definitions/mdpi} 

\usepackage[disable]{todonotes}
\usepackage{subcaption}

\firstpage{1} 
\pubvolume{1}
\issuenum{1}
\articlenumber{0}
\pubyear{2023}
\copyrightyear{2023}
\datereceived{ } 
\daterevised{ } 
\dateaccepted{ } 
\datepublished{ } 
\hreflink{https://doi.org/} 
\newcommand{\spcend}{\ensuremath{\:}}

\newcommand{\vect}[1]{\ensuremath{\mbox{\boldmath ${#1}$}}}

\newcommand{\likelihood}{\ensuremath{\mathcal{L}}}
\newcommand{\prior}{\ensuremath{\pi}}

\newcommand{\dx}{\ensuremath{\mathrm{\,d}}}

\newcommand{\temp}{\ensuremath{T_0}}

\newcommand{\prob}{\ensuremath{\text{P}}}
\newcommand{\given}{\ensuremath{{\,|\,}}}
\newcommand{\evidence}{\ensuremath{{z}}}

\newcommand{\data}{\ensuremath{\vect{y}}}

\renewcommand{\prob}{\ensuremath{{p}}}
\renewcommand{\vect}[1]{\ensuremath{{#1}}}
\renewcommand{\spcend}{\ensuremath{{}}}
\renewcommand{\evidence}{\ensuremath{{\mathcal{Z}}}}

\Title{Proximal nested sampling with data-driven priors \\for physical scientists}

\TitleCitation{Proximal nested sampling with data-driven priors for physical scientists}

\Author{Jason D.~McEwen $^{1,2}$*, Tobías I. Liaudat $^{1,3}$, Matthew A.~Price$^{1}$, Xiaohao Cai $^{4}$ and Marcelo Pereyra $^{5}$}

\AuthorNames{Jason D.~McEwen, Tobias I.~Liaudat, Matthew A.~Price, Xiaohao Cai and Marcelo Pereyra}

\AuthorCitation{McEwen, J.~D.; Liaudat, T.; Price, M.~A.; Cai, X.; Pereyra, M.}

\address{%
$^{1}$ \quad Mullard Space Science Laboratory, University College London (UCL), Dorking, RH5 6NT, UK;\\
$^{2}$ \quad Alan Turing Institute, London, NW1 2DB, UK;\\
$^{3}$ \quad Department of Computer Science, University College London (UCL), London, WC1E 6BT, UK;\\
$^{4}$ \quad School of Electronics and Computer Science, University of Southampton, Southampton, SO17 1BJ, UK;\\
$^{5}$ \quad School of Mathematical and Computer Sciences, Heriot-Watt University, Edinburgh, EH14 4AS, UK;}

\corres{Correspondence: jason.mcewen@gmail.com}

\abstract{Proximal nested sampling was introduced recently to open up Bayesian model selection for high-dimensional problems such as computational imaging.  The framework is suitable for models with a log-convex likelihood, which are ubiquitous in the imaging sciences. The purpose of this article is two-fold.  First, we review proximal nested sampling in a pedagogical manner in an attempt to elucidate the framework for physical scientists.  Second, we show how proximal nested sampling can be extended in an empirical Bayes setting to support data-driven priors, such as deep neural networks learned from training data.}


\begin{document}



\section{Introduction}

In much of the sciences not only is one interested in estimating the parameters of an underlying model, but deciding which model is best among a number of alternatives is of critical scientific interest. Bayesian model comparison provides a principled approach to model selection \cite{R01} that has found widespread application in the sciences \cite{ashton2022nested}.

Bayesian model comparison requires computation of the model evidence:
\begin{equation}
  \evidence =
  \prob(\data \given M)
  = \int \dx x \:
  \prob(\data \given x, M) \prob(x \given M)
  = \int \dx \vect{x} \:
  \likelihood(\vect{x}) \: \prior(\vect{x})
  \spcend ,
\end{equation}
also called the marginal likelihood, where $y \in \mathbb{R}^m$ denotes data, $x \in \mathbb{R}^n$  parameters of interest, and $M$ the model under consideration.  We adopt the shorthand notation for the likelihood of $\likelihood(\vect{x}) = \prob(\data \given x, M)$ and prior of $\prior(\vect{x}) = \prob(x \given M)$.  Evaluating the multi-dimensional integral of the model evidence is computationally challenging, particularly in high dimensions.
While a number of highly successful approaches to computing the model evidence have been developed, such as nested sampling \cite[\textit{e.g.}][]{S06, MPL06, FH08, FHB09, HHL15, ashton2022nested, buchner2021nested} and the learned harmonic mean estimator \cite{mce20, spurio-mancini:harmonic_sbi,polanska:harmonic_maxent}\todo{Need to update arxiv number of normalizing flow harmonic paper when have it.}, previous approaches do not scale to the very high-dimensional settings of computational imaging, which is our driving motivation.

The proximal nested sampling framework was introduced recently by a number of authors of the current article in order to open up Bayesian model selection for high-dimensional imaging problems \cite{cai:proximal_nested_sampling}.  Proximal nested sampling is suitable for models for which the likelihood is log-convex, which are ubiquitous in the imaging sciences.  By restricting the class of models considered, it is possible to exploit structure of the problem to enable computation in very high-dimensional settings of $\mathcal{O}(10^6)$ and beyond.

Proximal nested sampling draws heavily on convex analysis and proximal calculus.  In this article we present a pedagogical review of proximal nested sampling, sacrificing some mathematical rigor in an attempt to provide greater accessibility.  We also provide a concise review of convexity and proximal calculus to introduce the background underpinning the framework.  We assume the reader is familiar with nested sampling, hence we avoid repeating an introduction to nested sampling and instead refer the reader to other sources that provide excellent descriptions \cite{S06, ashton2022nested, buchner2021nested}.  Finally, for the first time we show in an empirical Bayes setting how proximal nested sampling can be extended to support data-driven priors, such as deep neural networks learned from training data.


\section{Convexity and proximal calculus}

We present a concise review of convexity and proximal calculus to introduce the background underpinning proximal nested sampling to make it more accessible.

\subsection{Convexity}

\begin{figure}[H]
  \vspace{-3mm}
  \subcaptionbox{Convex set}{\includegraphics[clip=true, trim={0 0 270 40}, width=0.22\textwidth]{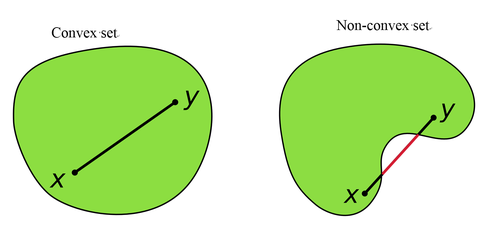}\vspace{5mm}}
  \subcaptionbox{Non-convex set}{\includegraphics[clip=true, trim={270 0 0 40}, width=0.22\textwidth]{figures/convex_sets.png}\vspace{5mm}}
  \subcaptionbox{Convex function}{\includegraphics[clip=true, trim={0 50 420 0}, width=0.25\textwidth]{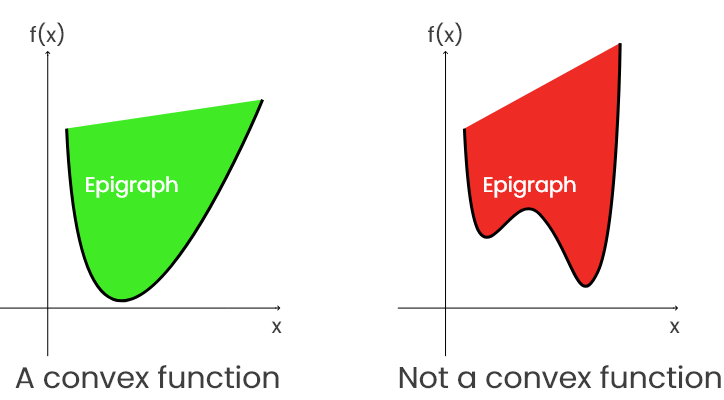}}
  \subcaptionbox{Non-convex function}{\includegraphics[clip=true, trim={390 50 30 0}, width=0.25\textwidth]{figures/epigraph.png}}
  \caption{Proximal nested sampling considers likelihoods that are log-convex and lower semicontinuous.  A lower semicontinuous convex function has a convex and closed epigraph.\label{fig:convexity}}
\end{figure}

Proximal nested sampling draws on convexity, key concepts of which are illustrated in Figure~\ref{fig:convexity}.  A set $\mathcal{C}$ is {convex} if for any $\vect{x}_1, \vect{x}_2 \in \mathcal{C}$ and $\alpha \in (0, 1)$ we have $\alpha \vect{x}_1 + (1 - \alpha) \vect{x}_2 \in \mathcal{C}$.The {epigraph} of a function $f : \mathbb{R}^n \rightarrow \mathbb{R}$ is defined by $\text{epi}(f )= \{ (\vect{x}, \gamma) \in \mathbb{R}^n \times \mathbb{R} \given f(\vect{x}) \leq \gamma \}$.  The function $f$ is {convex} if and only if its {epigraph is convex}.  A convex function is lower semicontinuous if its epigraph is closed (\textit{i.e.} includes its boundary).

\subsection{Proximity operator}

\begin{figure}
  \includegraphics[width=0.36\textwidth]{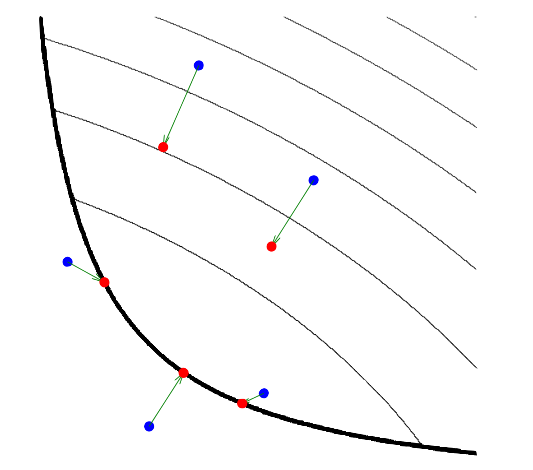}
  \caption{Illustration of the proximity operator (reproduced from \cite{NS13}). The proximal operator maps the blue points to red points (\textit{i.e.} from base to head of arrows).  The thick black line defines the domain boundary, while the thin black lines define level-sets (iso-contours) of $f$.  The proximity operator maps points towards the minimum of $f$, while remaining in the proximity of the original point. \label{fig:prox}}
\end{figure}

Proximal nested sampling leverages proximal calculus \cite{CP10,NS13}, a key component of which is the proximity operator, or prox.  The proximity operator of the function $f$ with parameter $\lambda$ is defined by
\begin{equation} \label{eqn:prox}
  \text{prox}_f^{\lambda} ({\vect x}) = \underset{\vect{u}}{\arg \min}
  \bigl[
  f({\vect u})
  +  \|{\vect u} - {\vect x}\|^2/2\lambda \bigr] .
\end{equation}
The proximity operator maps a point $\vect{x}$ towards the minimum of $f$, while remaining in the proximity of the original point.  The parameter $\lambda$ controls how close the mapped point remains to $\vect{x}$.  An illustration is given in Figure~\ref{fig:prox}.

The proximity operator can be considered as a generalisation of the projection onto a convex set.  Indeed, the projection operator can be expressed as a prox by
\begin{equation}
  \Pi_\mathcal{C} ({\vect x}) =
  \underset{\vect{u}}{\arg \min}
  \bigl [
  \chi_\mathcal{C}({\vect u})
  + \|{\vect u} - {\vect x}\|^2/2
  \bigr ]
  ,
\end{equation}
with function $f$ given by the characteristic function
$\chi_\mathcal{C}(\vect{x}) = \infty$ if $\vect{x} \notin \mathcal{C}$ and zero otherwise.

\subsection{Moreau-Yosida regularisation}

The final component required in the development of proximal nested sampling is Moreau-Yosida regularisation \cite[\textit{e.g.}][]{NS13}. The Moreau-Yosida envelop of a convex function $f : \mathbb{R}^n \rightarrow \mathbb{R}$ is given by the infimal convolution:
\begin{equation} \label{eqn:moreau-yosida}
  f^{\lambda}({\vect x}) = \inf_{{\vect u}\in \mathbb{R}^N}
  f({\vect u}) + \frac{\|{\vect u}- {\vect x}\|^2}{2\lambda} .
\end{equation}
The Moreau-Yoshida envelope of a function can be interpreted as taking its convex conjugate, adding regularisation, before taking the conjugate again \cite{NS13}.  Consequently, it provides a smooth regularised approximation of $f$, which is very useful to enable the use of gradient-based computational algorithms \cite[\textit{e.g.}][]{M15}.

The Moreau-Yosida envelop exhibits the following properties.
First, $\lambda$ controls the degree of regularisation with $f^{\lambda}({\vect x}) \rightarrow f({\vect x})$ as $\lambda \rightarrow 0$.
Second, the gradient of the Moreau-Yosida envelope of $f$ can be computed through its prox by $\nabla f^{\lambda}({\vect x}) = (\vect{x} - \text{prox}_f^{\lambda} ({\vect x}))/\lambda$.


\section{Proximal nested sampling}

The challenge of nested sampling in high-dimensional settings is to sample from the prior distribution subject to a hard likelihood constraint \cite{S06, ashton2022nested, buchner2021nested}.  Proximal nested sampling addresses this challenge for the case of log-convex likelihoods, which are widespread in computational imaging problems.  In this section we review the proximal nested sampling framework \cite{cai:proximal_nested_sampling} in a pedagogical manner, sacrificing some mathematical rigor in an attempt to improve readability and accessibility.

\subsection{Constrained sampling formulation}

Consider a prior and likelihood $\pi(x) \propto \text{exp}(-f(x))$ and ${\cal L}(x) \propto \text{exp}(-g(x))$, where the log-likehood $g=-\log\mathcal{L}$ is a convex lower semicontinuous function. The log-prior $f=-\log \pi$ need only be differentiable or convex (it need not be convex if it is differentiable).

We consider sampling from the prior $\pi({x})$, such that ${\cal L}(x) > L^*$ for some likelihood value $L^* \geq 0$. Let $\iota_{{L}^*}(x)$ and $\chi_{{L}^*}(x)$ be the indicator function	and characteristic function corresponding to this constraint, respectively, defined as
\begin{equation}
  \iota_{{L}^*}(x) =
  \begin{cases}
    1, & {\cal L} (x) > {L}^*, \\
    0, & \text{otherwise},
  \end{cases}
  \quad \text{and} \quad
  \chi_{{L}^*}(x) =
  \begin{cases}
    0,       & {\cal L} (x) > {L}^*, \\
    +\infty, & \text{otherwise}.
  \end{cases}
\end{equation}
Since $\log$ is monotonic,  ${\cal L} (x) > {L}^*$ is equivalent to
$g(x) < \tau$ for $\tau = -\log{L}^*$.
%
Explicitly define the convex set of the likelihood constraint by ${\cal B}_{\tau} = \{x \ \vert \ g(x) < \tau\}$. Then $\chi_{{L}^*}(x)$ is equivalent to $\chi_{{\cal B}_{\tau}}(x)$, where $\chi_{{\cal B}_{\tau}}(x) = \infty$ if $x \notin {\cal B}_{\tau}$ and zero otherwise.

Let ${\pi}_{{L}^*} (x) = \pi(x) \iota_{{L}^*}(x)$ represent the prior distribution with the hard likelihood constraint
${\cal L} (x) > {L}^*$. Since $\iota_{{L}^*}(x) = \text{exp}(- \chi_{{L}^*}(x))$, then we have
\begin{equation} \label{eqn:log-prior-cons}
  -\log \pi_{{L}^*} (x) = -\log\pi(x) + \chi_{{\cal B}_{\tau}}(x).
\end{equation}
To sample from the constrained prior we require sampling techniques that firstly can scale to high-dimensional settings and that secondly can support the convex constraint $\chi_{{\cal B}_{\tau}}(x)$.

\subsection{Langevin MCMC sampling}

Langevin Markov chain Monte Carlo (MCMC) sampling has been demonstrated to be highly effective at sampling in high-dimensional settings by exploiting gradient information \cite{M15,DMP16}.  The Langevin stochastic differential equation associated with distribution $p(x)$ is a stochastic process defined by
\begin{equation}
  {\rm d}{\vect{x}}(t)
  = \frac{1}{2}
  \nabla \log  p\bigl({\vect{x}}(t)\bigr)
  {\rm d}t
  + {\rm d} \vect{w}(t), 
\end{equation}
where $\vect{w}(t)$ is Brownian motion.  This process converges to $p(x)$ as time $t$ increases and is therefore useful for generating samples from $p(x)$.  In practice we compute a discrete-time approximation of $\vect{x}(t)$ by the conventional Euler-Maruyama discretisation:
\begin{equation}\label{eqn:eulerMaruyDisc}
  x^{(k+1)} = x^{(k)}
  + \frac{\delta}{2} \nabla \log p(x^{(k)})
  + \sqrt{\delta} w^{(k+1)},
\end{equation}
where $w^{(k)}$ is a sequence of standard Gaussian random variables and $\delta$ is a step size.

Equation~\ref{eqn:eulerMaruyDisc} provides a strategy for sampling in high-dimensions.  However, notice that the updates rely on the score of the target distribution $\nabla \log p(\cdot)$.  Nominally the target distribution must therefore be differentiable, which is not the case for our target of interest given by Equation~\ref{eqn:log-prior-cons}. The prior may or may not be differentiable but the likelihood constraint certainly is not.  Proximal versions of Langevin sampling have been developed to address the setting where the distribution is log-convex but not necessarily differentiable \cite{M15,DMP16}.  We follow a similar approach.

\subsection{Proximal nested sampling framework}

The proximal nested sampling framework follows by taking the constrained sampling formulation of Equation~\ref{eqn:log-prior-cons}, adopting Langevin MCMC sampling of Equation~\ref{eqn:eulerMaruyDisc}, and applying Moreau-Yosida regularisation of Equation~\ref{eqn:moreau-yosida} to the convex constraint $\chi_{{\cal B}_{\tau}}(x)$ to yield a differentiable target.  This strategy yields (see \cite{cai:proximal_nested_sampling}) the update equation:
\begin{equation} \label{eqn:proxnest_update_1}
  \vect{x}^{(k+1)} =
  \vect{x}^{(k)}
  + \frac{\delta}{2}\nabla \log\pi(\vect{x}^{(k)})
  - \frac{\delta}{2\lambda}
  \bigl[ \vect{x}^{(k)} - \text{prox}_{\chi_{{\cal B}_{\tau}}}(\vect{x}^{(k)}) \bigr]
  + \sqrt{\delta} {\vect{w}}^{(k+1)},
\end{equation}
where $\delta$ is the step size and $\lambda$ is the Moreau-Yosida regularisation parameter.

\begin{figure}
  \includegraphics[clip=true, trim={0 10 0 10}, width=0.51\textwidth]{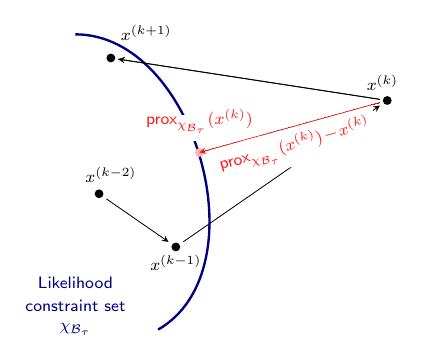}
  \caption{Diagram illustrating proximal nested sampling.  If a sample $x^{(k)}$ outside of the likelihood constraint is considered, then proximal nested sampling introduces a term in the direction of the projection of $x^{(k)}$ onto the convex set defining the likelihood constraint, thereby acting to push the Markov chain back into the constraint set ${\cal B}_\tau$ if it wanders outside of it.  A subsequent Metropolis-Hastings step can be introduced to enforce strict adherence to the convex likelihood constraint. \label{fig:proxnest}}
\end{figure}

Further intuition regarding proximal nested sampling can be gained by examining the term $v^{(k)} = -[ \vect{x}^{(k)} - \text{prox}_{\chi_{{\cal B}_{\tau}}}(\vect{x}^{(k)}) ]$, together with Figure~\ref{fig:proxnest}.
The vector $v^{(k)}$ points from the sample ${x}^{(k)}$ to its projection onto the likelihood constraint.
If the sample $x^{(k)}$ is already in the likelihood-restricted prior support ${\cal B}_\tau$, \textit{i.e.} $x \in {\cal B}_\tau$, the term \mbox{$v^{(k)}$} disappears and the Markov chain iteration simply involves the standard Langevin MCMC update.  In contrast, if  $x^{(k)}$ is not in ${\cal B}_\tau$, \textit{i.e.} $x \notin {\cal B}_\tau$, then a step is taken in the direction \mbox{$v^{(k)}$}, which acts to move the next iteration of the Markov chain in the direction of the projection of $x^{(k)}$ onto the convex set ${\cal B}_\tau$. This term therefore acts to push the Markov chain back into the constraint set ${\cal B}_\tau$ if it wanders outside of it.\footnote{Note that proximal nested sampling has some similarity with Galilean \cite{skilling2012bayesian} and constrained Hamiltonian \cite{betancourt2011nested} nested sampling.  In these approaches Makov chains are also considered and if the Markov chain steps outside of the likelihood-constraint then it is reflected by an approximation of the shape of the boundary.}

We have so far assumed that the (log) prior is differentiable (see Equation~\ref{eqn:proxnest_update_1}).  This may not be the case, as is typical for sparsity-promoting priors (\textit{e.g.} $-\log\pi(\vect{x}) = \|\Psi^\dagger \vect{x} \|_1 + \text{const.}$ for some wavelet dictionary $\Psi$).  Then we make a Moreau-Yosida approximation of the log-prior, yielding the update equation:
\begin{equation}
  \vect{x}^{(k+1)} =
  \vect{x}^{(k)}
  - \frac{\delta}{2\lambda} \bigl[ \vect{x}^{(k)} - \text{prox}_{-\log\pi}^\lambda(\vect{x}^{(k)}) \bigr]
  - \frac{\delta}{2\lambda} \bigl[ \vect{x}^{(k)} - \text{prox}_{\chi_{{\cal B}_{\tau}}}(\vect{x}^{(k)}) \bigr]
  + \sqrt{\delta} {\vect{w}}^{(k+1)}
  \spcend .
  \label{eq:prox_smpling}
\end{equation}
For notational simplicity here we have adopted the same regularisation parameter $\lambda$ for each Moreau-Yosida approximation.

With the current formulation we are not guaranteed to recover samples from the prior subject to the hard likelihood constraint due to the approximation introduced in the Moreu-Yosida regularisation and due to the approximation in discretising the underlying Langevin stochastic differential equation.  We therefore introduce a Metropolis-Hastings correction step to eliminate the bias introduced by these approximations and ensure convergence to the required target density (see \cite{cai:proximal_nested_sampling} for further details).

Finally, we adopt this strategy for sampling from the constrained prior in the standard nested sampling strategy to recover the proximal nested sampling framework.  The algorithm can be initalised with samples from the prior as described by the update equations above but with the likelihood term removed, \textit{i.e.} with $[ \vect{x}^{(k)} - \text{prox}_{\chi_{{\cal B}_{\tau}}}(\vect{x}^{(k)}) ] \rightarrow 0$.

\subsection{Explicit forms of proximal nested sampling}
\label{sc:explicit_forms}

While we have discussed the general framework for proximal nested sampling, we have yet to address the issue of computing the proximity operators involved.  As Equation~\ref{eqn:prox} demonstrates, computing proximity operators involves solving an optimisation problem.  Only in certain cases are closed form solutions available \cite{CP10}.  Explicit forms of proximal nested sampling must therefore be considered for the problem at hand.

We focus on a common high-dimensional inverse imaging problem where we acquire noisy observations $y = \Phi x + n$, of an underlying image $x$ via some measurement model $\Phi$, in the presence of Gaussian noise $n$ (without loss of generality we consider independent and identically distributed noise here).  We consider a Gaussian negative likelihood, $-\log\mathcal{L}(\vect{x}) = \bigl\| \vect{y} - \vect{{\Phi}} \vect{x} \bigr\|_2^2/2\sigma^2 + \text{const.}$, and a sparsity-promoting prior, $-\log \pi(\vect{x}) = \mu \bigl\|\Psi^\dagger \vect{x}\bigr\|_1 + \text{const.}$, for some wavelet dictionary $\Psi$.  The prox of the prior can be computed in closed-form by \cite{CP10}
\begin{equation}
  \text{prox}_{-\log\pi}^\lambda(\vect{x}) = \vect{x} + \Psi \bigl(\text{soft}_{\lambda \mu}(\Psi^{\dagger} \vect{x}^\prime) -\Psi^{\dagger} \vect{x} \bigr),
  \label{eq:sparsity_prox}
\end{equation}
where $\text{soft}_\lambda(\cdot)$ is the soft thresholding function with threshold $\lambda$ (recall $\mu$ is the scale of the sparsity-promoting prior, \textit{i.e.} the regularisation parameter, defined above).
However, the prox of the likelihood is not so straightforward.  The prox for the likelihood can be recast as a saddle-point problem
that can be solved iteratively by a primal dual method initialised by the current sample position (see \cite{cai:proximal_nested_sampling} for further details):
\begin{enumerate} \small
  \item $
          z^{(i+1)}
          = z^{(i)} + \delta_1\Phi \bar{x}^{(i)} - \text{prox}_{\chi_{{\cal B}^\prime_{\tau^\prime }}} (z^{(i)} + \delta_1\Phi \bar{x}^{(i)}),
        $
        \begin{equation*} \hspace{-25mm}
          \text{where }\text{prox}_{\chi_{{\cal B}^\prime_{\tau^\prime }}} (z) =
          \text{proj}_{{{\cal B}^\prime_{\tau^\prime }}} (z) =
          \begin{cases}
            z,                                                      & \text{if} \  z \in {\cal B}^\prime_{\tau^\prime }, \\
            \frac{z - y }{\|z - y \|_2} \sqrt{2 \tau \sigma^2} + y, & \text{otherwise};
          \end{cases}
        \end{equation*}
  \item $x^{(i+1)} =  (x^\prime + x^{(i)} - \delta_2{\Phi}^\dagger z^{(i+1)})/2$ ;
  \item $\bar{x}^{(i+1)} = x^{(i+1)} + \delta_3 (x^{(i+1)} - x^{(i)})$ .

\end{enumerate}

Combining these algorithms to efficiently compute prox operators with the proximal nested sampling framework, we can compute the model evidence to perform Bayesian model comparison in high-dimensional settings.  We can also obtain posterior distributions with the usual weighted samples from the dead points of nested sampling.  This allows one to recover, for example, point estimates such as the posterior mean image.


\section{Deep data-driven priors}

While hand-crafted priors, such as wavelet-based sparsity promoting priors, are common in computational imaging, they provide only limited expressivity.  If example images are available an empirical Bayes approach with data-driven priors can be taken, where the prior is learned from training data.  Since proximal nested sampling requires only the log-likelihood to be convex, complex data-driven priors, such as represented by deep neural networks, can be integrated into the framework.  Through Tweedie's formula we describe how proximal nested sampling can be adapted to support data-driven priors, opening up Bayesian model selection for data-driven approaches.  We take a similar approach to \cite{laumont2022bayesian}, where data-driven priors are integrated into Langevin MCMC sampling strategies, although in that work model selection is not considered.

\subsection{Tweedie's formula and data-driven priors}

Tweedie's formula is a remarkable result in Bayesian estimation credited to personal correspondence with Maurice Kenneth Tweedie \cite{robbins:1956}.  Tweedie's formula has gained renewed interest in recent years \cite{efron2011tweedie,laumont2022bayesian,kim2021noise2score,chung2022improving} due to its connection to score matching \cite{sohl2015deep,song2019generative,song2020improved} and denoising diffusion models \cite{song2020score,rombach2022high}, which as of this writing provide state-of-the-art performance in deep generative modelling.

Tweedie's result follows by considering the following scenario.  Consider $x$ sampled from a prior distribution $q(\cdot)$ and noisy observations $z \sim \mathcal{N}(x, \sigma^2 I)$.  Tweedie's formula gives the posterior expectation of $x$ given $z$ as
\begin{equation}
  \label{eq:tweedie}
  E(x \vert z) = z + \sigma^2 \nabla \log p(z) ,
\end{equation}
where $p(z)$ is the marginal distribution of $z$ (for further details see, \textit{e.g.}, \cite{efron2011tweedie}).  The critical advantage of Tweedie's formula is that it does not require knowledge of the underlying distribution $q(\cdot)$ but rather only the marginalised distribution of the observation.  Equation~\ref{eq:tweedie} can be interpreted as a denoising strategy to estimate $x$ from noisy observations $z$.  Moreover, Tweedie's formula can also be used to relate a denoiser (potentially a trained deep neural network) to the score $\nabla \log p(z)$.

In a data-driven setting, where the underlying prior is implicitly specified by training data (which are considered to be samples from the prior), there is no guarantee that the underlying prior, and therefore the posterior, is well-suited for gradient-based Bayesian computation such as Langevin sampling, \textit{e.g.} it may not be differentiable.  Therefore we consider a regularised version of the prior defined by Gaussian smoothing:
\begin{equation}
  p_\epsilon(x) = (2\pi \epsilon)^{-n/2} \int \text{d}x^\prime \text{exp}( |\ x - x^\prime\|^2_2 / (2\epsilon))  q(x^\prime) .
\end{equation}

This regularisation can also be viewed as adding a small amount of regularising Gaussian noise.  We can therefore leverage Tweedie's formula to relate the regularised prior distribution $p_\epsilon(x)$ to a denoiser $D_\epsilon$ trained to recover $x$ from noisy observations $x_\epsilon \sim \mathcal{N}(x, \epsilon I)$, \textit{i.e.} the score of the regualised prior can be related to the denoiser by
\begin{equation}
  \nabla \log p_\epsilon(x) =  \epsilon^{-1} (D_\epsilon(x) - x).
\end{equation}

Denoisers are commonly integrated in proximal optimisation algorithms in replace of proximity operators, giving rise to so-called plug-and-play (PnP) methods \cite{venkatakrishnan2013plug, ryu2019plug}  and, more recently, also into Bayesian computational algorithms \cite{laumont2022bayesian}.  Typically denoisers are represented by deep neural networks, which can be trained by injecting a small amount of noise in training data and learning to denoise the corrupted data.  While a noise level $\epsilon$ needs to be chosen, as discussed above this is considered a regularisation of the prior and so the denoiser need not be trained on the noise level of a problem at hand.  In this manner, the same denoiser can be used for multiple subsequent problems (hence the PnP name).  The learned score of the regularised prior inherits the same properties as the denoiser, such as smoothness, hence the denoiser should be considered carefully.  Well-behaved denoisers have been considered already in PnP methods (in order to provide convergence guarantees) and a popular approach for imaging problems is the DnCNN model \cite{ryu2019plug}, which is based on a deep convolutional neural network, and that is (Lipschitz) continuous.

\subsection{Proximal nested sampling with data-driven priors}

By Tweedie's formula the standard proximal nested sampling update of Equation~\ref{eqn:proxnest_update_1} can be revised to integrate a learned denoiser, yielding
\begin{equation}
  x^{(k+1)} =
  x^{(k)}
  - \frac{\alpha \delta}{2\epsilon} \bigl[ x - D_\epsilon(x^{(k)}) \bigr]
  - \frac{\delta}{2\lambda} {\bigl[ x^{(k)} - \text{prox}_{\chi_{{\cal B}_{\tau}}}(x^{(k)}) \bigr]}
  + \sqrt{\delta} { w}^{(k+1)},
  \label{eq:data_driven_sampling}
\end{equation}
where we have included a regularisation parameter $\alpha$ that allows us to balance the influence of the prior and the data fidelity terms \citep{laumont2022bayesian}.
We typically consider a deep convolutional neural network based on the DnCNN model \cite{ryu2019plug} since it is (Lipschitz) continuous and has been demonstrated to perform very well in PnP settings \cite{ryu2019plug,laumont2022bayesian}.  Again, this sampling strategy can then be integrated into the standard nested sampling framework.

We can therefore support data-driven priors in the proximal nested sampling framework by integrating a deep denoiser that learns to denoise training data, using Tweedie's formula to relate this to the score of a regularised data-driven prior.


\section{Numerical experiments}

\begin{figure}
  \subcaptionbox{Ground truth}{\includegraphics[width=0.24\textwidth]{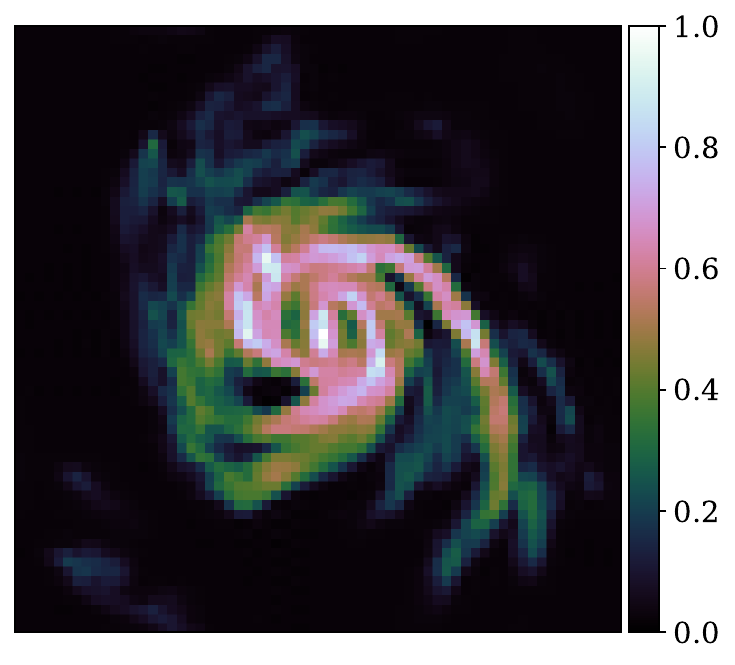}}
  \subcaptionbox{Dirty}{\includegraphics[width=0.24\textwidth]{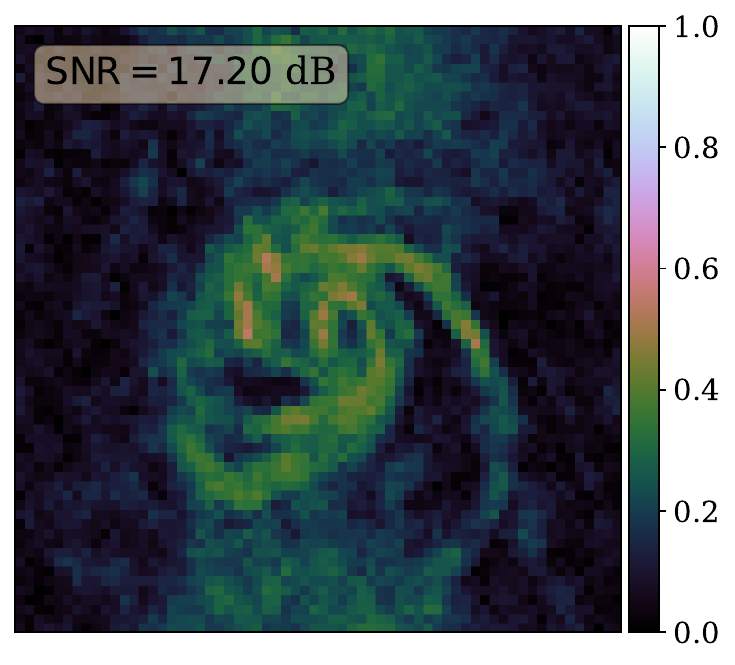}}
  \subcaptionbox{Hand-crafted prior}{\includegraphics[width=0.24\textwidth]{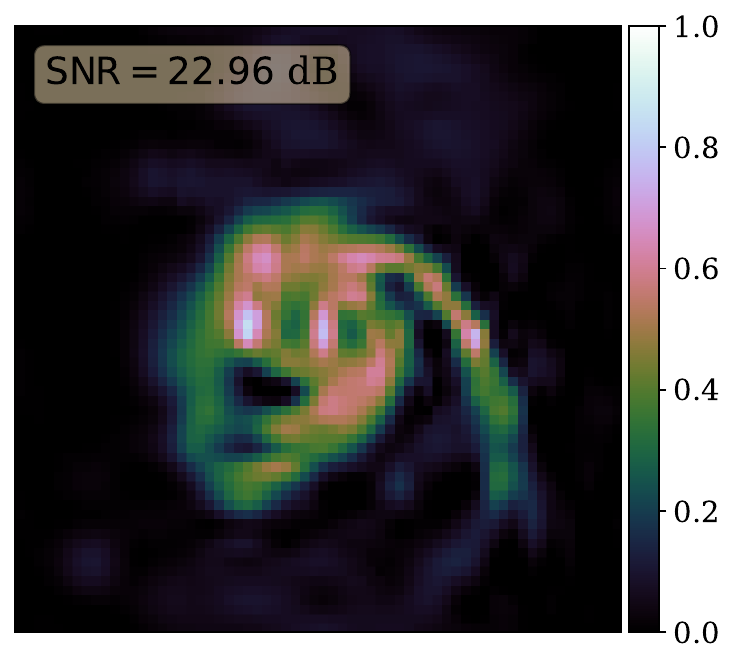}}
  \subcaptionbox{Data-driven prior}{\includegraphics[width=0.24\textwidth]{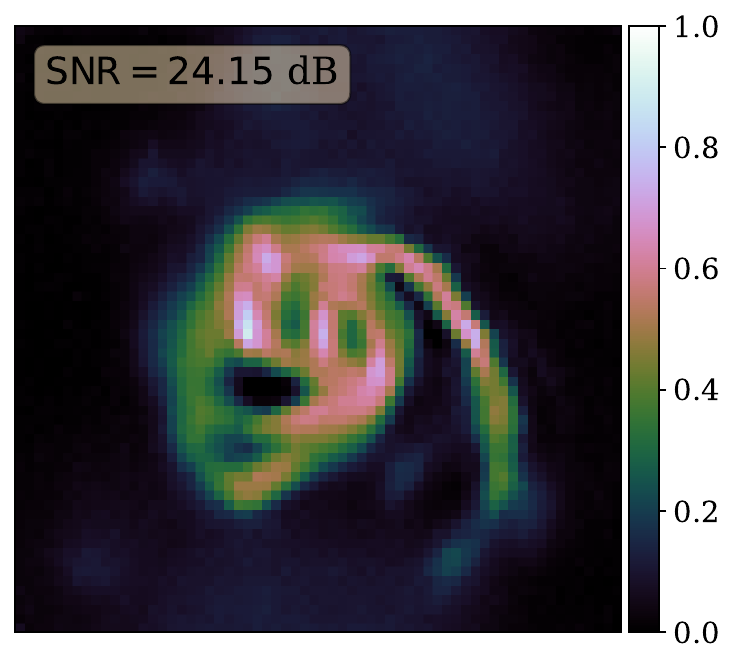}}
  \caption{Results of radio interferometric imaging reconstruction problem. \textbf{(a)} Ground truth galaxy image. \textbf{(b)} Dirty reconstruction based on pseudo-inverting the measurement operator $\Phi$. \textbf{(c)} Posterior mean reconstruction computed from proximal nested samples for the hand-crafted wavelet-sparsity prior. \textbf{(d)} Posterior mean reconstruction for the data-driven prior based on a deep neural network (DnCNN) trained on example images.  Reconstruction SNR is shown on each image.  The computed SNR levels demonstrate that the data-driven prior results in a superior reconstruction quality, although this may not be obvious from a visual assessment of the reconstructed images.  Computing the reconstructed SNR requires knowledge of the ground truth, which is not available in realistic settings. 	The Bayesian model evidence proves a way to compare the hand-crafted and data-driven models without requiring knowledge of the ground truth.  For this example the Bayesian evidence correctly selects the data-driven prior as the best model.
    \label{fig:img_results_radio}}
\end{figure}

The new methodology presented allows us to perform Bayesian model comparison between a data-driven and hand-crafted prior(validation of proximal nested sampling in a setting where the ground truth can be computed directly has been performed already \cite{cai:proximal_nested_sampling}). We consider a simple radio interferometric imaging reconstruction problem as an illustration.  We assume the same observational model as Section~\ref{sc:explicit_forms}, with white Gaussian noise giving a signal-to-noise ratio (SNR) of $15$dB. The measurement operator $\Phi$ is a masked Fourier transform as a simple model of a radio interferometric telescope. The mask is built by randomly selecting $50\%$ of the Fourier coefficients.
A Gaussian likelihood is used in both models. For the hand-crafted prior we consider a sparsity-promoting prior using a Daubechis 6 wavelet dictionary. We base the data-driven prior on a DnCNN \cite{ryu2019plug} model trained on galaxy images extracted from the IllustrisTNG simulations \cite{nelsonIllustrisTNGSimulationsPublic2019}. We also consider an IllustrisTNG galaxy simulation, not used in training, as the ground truth test image.
%
We generate samples following the proximal nested sampling strategies of Equation~\ref{eq:prox_smpling} and Equation~\ref{eq:data_driven_sampling} for the hand-crafted and data-driven priors, respectively. Posterior inferences (e.g.\ posterior mean image) and the model evidence can then be computed from nested sampling samples in the usual manner.
The step size $\delta$ is set to $10^{-7}$, the Moreau-Yosida regularisation parameter $\lambda$ to $5 \times 10^{-7}$, and the regularisation strength of wavelet-based model $\mu$ to $5 \times 10^{4}$. We consider noise level $\epsilon \simeq 8.34$ and set the regularisation parameter $\alpha$ of the data-driven prior to $3.5 \times 10^{-7}$. For the nested sampling methods, the number of live and dead samples is set to $10^{2}$ and $2.5 \times 10^{3}$, respectively. For the Langevin sampling, we use a thinning factor of $20$ and set the number of burn-in iterations to $10^{2}$.

Results are presented in Figure~\ref{fig:img_results_radio}.  The data-driven prior results in a superior reconstruction with an improvement in SNR of $1.2$dB, although it may be difficult to tell simply from visual inspection of the recovered images.  Computing the SNR of the reconstructed images requires knowledge of the ground truth, which clearly is not accessible in realistic settings involving real observational data.  The Bayesian model evidence, computed by proximal nested sampling, proves a way to compare the hand-crafted and data-driven models without requiring knowledge of the ground truth and is therefore applicable in realistic scenarios. We compute log evidences of $-2.96\times10^{3}$ for the hand-crafted prior and $-1.35\times10^{3}$ for the data-driven prior. Consequently, the data-driven model is preferred by the model evidence, which agrees with the SNR levels computed from the ground truth.  These results are all as one might expect since learned data-driven priors are more expressive than hand-crafted priors and can better adapt to model high-dimensional images.

\section{Conclusions}

Proximal nested sampling leverages proximal calculus to extend nested sampling to high-dimensional settings for problems involving log-convex likelihoods, which are ubiquitous in computational imaging.  The purpose of this article is two-fold.  First, we review proximal nested sampling in a pedagogical manner in an attempt to elucidate the framework for physical scientists.  Second, we show how proximal nested sampling can be extended in an empirical Bayes setting to support data-driven priors, such as deep neural networks learned from training data.  We show only preliminary results for learned proximal nested sampling and will present a more thorough study in a follow-up article.

\vspace{6pt}




\authorcontributions{Conceptualization, J.D.M. and M.P.; methodology, J.D.M., X.C. and M.P.; software, T.I.L., M.A.P. and X.C.; validation, T.I.L., M.A.P. and X.C.; resources, J.D.M.; data curation, M.A.P.; writing---original draft preparation, J.D.M. and T.I.L.; writing---review and editing, J.D.M., T.I.L., M.A.P., X.C. and M.P.; supervision, J.D.M.; project administration, J.D.M.; funding acquisition, J.D.M., M.A.P. and M.P. All authors have read and agreed to the published version of the manuscript.}

\funding{This research was funded by EPSRC grant number EP/W007673/1.}

\dataavailability{The \texttt{ProxNest} code and experiments are available at \url{https://github.com/astro-informatics/proxnest}.}




\conflictsofinterest{The authors declare no conflict of interest.}

\begin{adjustwidth}{-\extralength}{0cm}

	\reftitle{References}


	\bibliography{refs_xhcai,refs_proxnest_paper}

	\PublishersNote{}
\end{adjustwidth}
\end{document}